Imaging ferroelectric hysteresis and domain wall pinning

S. Bhattacharya [1] and M.J. Higgins [2]

[1] Tata Institute of Fundamental Research, Homi Bhabha Road, Mumbai – 400005, India

[2] Princeton High School, Princeton, New Jersey 08540, USA

Simultaneous imaging of the piezoresponse phase, amplitude and bare surface topography of displacive ferroelectric thin films by scanning probe microscopy directly shows the nature of domain wall pinning and its relation to morphological disorder. Strong and stable pinning of walls occurs at grain boundaries while weak and unstable pinning occurs within the grains. The results show that polarization reversal consists of both nucleation of domains and spreading of domain walls, the latter being the dominant process in multi-grain regions. A characteristic "single-grain" switching in faceted grains is observed as the motion of a single domain wall parallel to one of the facets.

An understanding of domain wall dynamics associated with hysteresis phenomena in condensed matter ground states such as a ferroelectric or a ferromagnet, is important in the general area of statistical physics of elastic media in a random pinning environment [1]. The generically *metastable* equilibrium behavior of such systems and their non-equilibrium evolution under an external driving force, leading to hysteresis, are of current interest because of their relevance in a variety of condensed matter systems such as vortices in superconductors [2], sliding charge density waves [3], electron lattices in semiconductor heterostructures [4], two-fluid interfaces in random porous media [5], in addition to the general problem of sliding friction [6].

Furthermore, ferroelectricity, like ferromagnetism, forms the basis of non-volatile memory through the hysteresis phenomenon where the spontaneous polarization of a polarized ferroelectric is retained even after the polarizing field is turned off [7]. The utility of these materials as memory depends critically on their ability to (a) switch polarization repeatedly and robustly from one state to another, (say "up" to " down") at suitably small polarizing fields and (b) retain this polarization against thermal or other randomizing influences. Therefore, an understanding of the dynamics of domain walls [8] is central to both the basic physics involved and the principles for designing robust memory. Historically, these phenomena were studied through bulk measurements, whose interpretation required assumptions not easy to ascertain *apriori.* Recently, however, more direct scanning probe imaging studies in the nanoscale have begun to provide incisive information about the complex process leading to repeatable hysteresis, or the absence thereof [9].

In this letter we report on a scanning probe microscopy-based study of a ferroelectric thin film that focuses on a central issue, namely, the nature of domain wall pinning. By *simultaneous imaging* of the magnitude and direction of the polarization, together with the topography, we probe the stability of domain walls, the types of pinning and their connection to quenched morphological disorder. A broad characterization of pinning, weak or strong, random or correlated, point-like or extended, elucidates connections with other condensed matter ground states mentioned above and to the theoretical problem of disordered elastic media, in general. All measurements reported here are on c-axis oriented thin films of lead-zirconate/ titanate (PZT) with Zr/Ti ratio of 20/80 of typical thickness of 50 nm, grown by sol-gel method, on a platinum bottom electrode grown on Si-substrates. The imaging studies were made with a Digital 3000 atomic force microscope suitably modified to perform piezo-response studies using metallic tips and an external but standard phase-sensitive lock-in detection assembly.

Piezo-response Force Microscopy (PFM) is a standard method [10] of visualizing ferroelectric domains of different polarizations. The ferroelectric moment is proportional to the piezoelectric coefficient; thus for a c-axis oriented sample with an effective electric field also along the c-axis, the technique measures the piezoelectric coefficient $d_{33}$. The strain measured is given by $\delta \sim \int d_{33}(z)E_3dz$ (where $E_3$ is an oscillatory electric field of frequency 11kHz in the experiment), integrated over the effective thickness of the sample. The order parameter ***P*** is then proportional to $|R| \exp(i\,\phi)$ where $|R|$, the amplitude of the resulting oscillatory strain, is a continuous variable, summed over the thickness of the sample. But the phase $\phi$ is discrete, 0 and $\pi$, for the so-called $c^+$ and $c^-$ domains, corresponding to up or down polarization. In our experiments, the metallic tip of the AFM was used to "write" or polarize the sample using a voltage $V_{dc}$, which is greater than the local coercive voltage $V_c$. The tip is rastered along a predetermined path to generate the desired "writing". This written pattern is then read by using the PFM mode with an ac reading voltage $V_{ac}$ significantly smaller than $V_c$. Typical results of such writing and reading are shown in Figure 1 where the topography, amplitude and phase, measured simultaneously, are plotted in panels a, b and c respectively. A square region was first poled by the tip to "down" polarization. A "cross" was then written in the center, with the "up" polarization and the written pattern was "read" using the piezo-response microscopy method. The amplitude signal is large inside the domains regardless of the sign of the polarization but is small at the boundary representing the domain wall (DW), across which the polarization switches, i.e., the dark lines are convenient markers of the DW, consistent with previous studies [11].

The *unintended jaggedness* of the domain wall is an indicator of the quality of "writing". It limits the density of memory elements and is of central importance in memory design. The physical origin of this limitation is illustrated in panel d which is generated by a weighted sum of the raw signals in the amplitude and topography images in panels (a) and (b). As is obvious in panel (d), the boundaries of the written pattern closely follow the contour of contiguous grain boundaries (GB).

Importantly, the DW's appear in two varieties, sharp and fuzzy. The sharp walls correspond precisely to the GB's. The width of the sharp walls is typically about 20nm, which is determined from independent estimates to be the effective size of the tip and does not represent the intrinsic width of the DW's, which is likely to be significantly smaller [10]. However, it shows unambiguously that the GB's, which are extended defects, are the favored pinning sites for the domain wall. This is easily understood: since the ferroelectric order parameter is strongly coupled to lattice distortion, the GB is where the order parameter is suppressed and is thus energetically the best place to locate the DW. We note that the c-axis oriented films have a columnar morphology and thus the GB's represent extended columnar defects, analogous to columnar pins in vortex phases of superconductors [12]. The image in (b) is not bi-modal as in (c) and shows a large variation of the magnitude of R within regions with the same phase $\phi$ as in the upper panel. Even so, the amplitude is remarkably uniform within a grain but shows large grain-to-grain variation for polarization of the same sign. Thus, the correlation length $\xi_R$ for the continuous variable R is typically given by the grain size, while the correlation length $\xi_\phi$ for the discrete variable $\phi$ can be much larger, spanning multi-grain regions in the poled state.

The fuzzy domain walls, in contrast with the sharp walls, are located in the intra-granular space; two of them are encircled on each panel in Figure 1. They are very wide, covering almost the entire grain, nearly an order of magnitude larger (~100nm) than the sharp walls and are not limited by the tip-size. *These fuzzy walls are characterized by (1) phase signal varying randomly by $\pi$ from one pixel to the next and (2) by a large fluctuation in the overall small amplitude signal within these wall areas*. A more detailed analysis shows that a significant source of the fuzziness is temporal in nature, i.e. the signal is temporally noisy, the details of which will be reported elsewhere [13]. But the fuzziness clearly implies "weak" pinning, (presumably due to random point-defects such as oxygen vacancies or substitutional impurities) in the intra-granular space.

Using the same methods, we have imaged the switching of a multigrain region, as described below. A specific area is first written using a dc bias voltage $V_{dc}$ by the tip. As the tip moves over to the next area to be written, the previously written area relaxes towards its remanent state ($V_{dc} = 0$). Once the entire area is thus written on and then allowed to relax, we read the resulting domain structure with the voltage applied to the tip is $V_{dc} + V_{ac}$ , where $V_{ac}$ is the piezo-detection voltage ($V_{ac} << V_{dc}$). This is equivalent to executing a "minor hysteresis loop" commonly used in ferromagnets: the state to be read is thus closer to the original hysteresis loop than to the remanent state.

Figure 2 shows a *composite* of phase and topography images of switching of a 500nmx500nm area. The upper left panel represents the remanent state after the area was poled "up" and then allowed to relax. Every subsequent panel marks the $V_{dc}$ value at which the state is "read". The images show a grain-by-grain spreading of the polarization reversal from up to down among contiguous grains, often preceded by a sliver of switched region extending along the grain boundary and then spreading through a grain. One such spreading front is marked by the black arrows for a grain enclosed by a dark square on the first panel. This process is analogous to the easy movement of vortices in oxide superconductors along twin boundaries but not across them [14]. Isolated nucleation also occurs [15], as shown in panel (c) by the encircled area, but is a secondary contribution to the net polarization.

Combining these results, we conclude that the switching process is dominated by DW-motion in c-axis oriented polycrystalline films that are realistic candidates for non-volatile memory. The DW is stable due to the strong pinning by the correlated disorder provided by the GBs. The intra-granular space, on the other hand, presents weak and presumably collective pinning that results in unstable DW's that are seldom stable enough to be captured easily in an image. They are mostly visible as fluctuating fuzzy DWs. The implication of these results is obvious: In systems such as these, the memory cells should be grain-size matched [16] for optimal stability and large polarization. The fundamental reason for this is the coupling of the polarization to lattice distortion that makes ferroelectricity extremely sensitive to extended morphological disorder such as the GBs.

Finally, we demonstrate switching observed in a single grain. The process of single grain switching is typically fast and occurs within a very narrow range of bias values so that at a given dc bias, the imaging technique captures few DWs in intra-granular space. However, if the grains are faceted, i.e., the GB's correspond to crystallographic axes, the evolution of the single grain switching spans a large enough bias range to be captured by these imaging techniques. A remarkable example is shown in Fig.3. In this case, a large square area was written in "up" polarization. But the central grain remained polarized in the "down" direction, implying a larger local coercivity. The relaxed state is found to be clearly faceted, as in epitaxial samples [17]; a few of the facets are marked by dotted lines in panel(a). The subsequent panels are obtained at higher dc biases aimed at switching the central grain back to the "up" polarization. The left and right panels show the amplitude and phase images, respectively. As $V_{dc}$ increases, we observe the remarkable movement of a single domain wall, marked by the black arrow in the amplitude panels, through the grain. Very significantly, DW moves in a way parallel to one of the facets, marked by the white arrow, as it sweeps across and becomes rounded only at the very late stages of switching. At the beginning of the switching process, the interior of the grain has large amplitude as do the neighboring grains. As $V_{dc}$ increases, the DW broadens as it moves towards the interior of the grain depleting the amplitude ahead of it and enhancing the amplitude behind, while the back wall, marked by the white arrow, remains pinned to the initial configuration. In the last panel (h) the back wall becomes rounded as the domain wall approaches collapse with the switching approaching completion. Similarly, the phase fluctuates in the wall region as seen in Fig1(c), but in contrast with Fig.1, the fluctuating walls are now stable enough to be captured reliably in the images.

The movement of the domain wall with increasing dc bias implies a large gradient in the local coercivity. The faceted grain boundaries, an analog of both twin boundaries and columnar defects in superconductors, provide the largest local coercivity, a measure of the local pinning strength, which can be as much as an order of magnitude larger than non-faceted ones. Moreover, the local coercivity varies significantly from one facet to another. As the dc bias is increased, the back wall remains pinned to the faceted GB, which presumably has the strongest pinning. The front wall moves through the intra-granular space where uncorrelated local point defects provide weaker pinning than the extended GB. This explains the motion of the domain wall but not the specific direction of motion, nearly parallel to the pinned back wall, nor their stability. We tentatively explain it in the following way. The back wall is the strongest pinning GB; the long-range elastic interaction is responsible for creating a decreasing gradient of local coercivity away from the back wall. The leading edge of the moving domain wall climbs this gradient as the bias is increased. Thus, the pinning in the GB's controls the DW-motion in the intra-granular space and accounts for its stability. We also note that the observed motion of the domain wall is remarkably similar to the motion of the boundary between an ordered and a disordered vortex phase in superconductors where the system has a large gradient in the "local critical current" (the analog of a local coercive field) between the edge of a sample and the interior [18].

The imaging results shown above provide a broad scenario of polarization-reversal in device-quality c-axis oriented ferroelectric films. Due to the strong coupling between the order parameter (ferroelectric moment) and the lattice distortion, lattice defects suppress the ferroelectric order parameter becoming natural pinning centers for the DW. The results yield the following conclusions. (I) Extended columnar defects such as the grain boundaries are strong pinning sites for the DW. (II) The switching is largely "collective" in the weak-pinning intragranular space that produces unstable and noisy domain walls, characterized by fuzzy images and are thus unsuitable for stable and nonvolatile memory. Although the phase correlation length can be larger depending on the poling field, the amplitude correlation length is typically set by the grain size. FE-based memory in polycrystalline material should thus be grain-size matched. (III) Switching of multi-grain areas occurs through spreading of domain walls across contiguous grains as the dominant contributor to the net polarization, whereas local nucleation of disconnected domains is a secondary process. (IV) Single-grain switching for faceted grains proceeds via the motion of a single DW. In this case the GB pinning due to long-range elastic interaction controls the DW dynamics in intra-granular space. These results provide a characterization of the ferroelectric DW motion in the same common conceptual framework of a classification and hierarchy of pinning for the dynamics of disordered elastic media as has been done for other closely related systems [1,2,3]. We note that similar jaggedness for both ferroelectric and ferromagnetic domain walls, correlated with morphological disorder, has recently been reported in multiferroic systems as well [19]. These results are thus of importance in memory design principles in a wider class of systems of current interest. Clearly, a comprehensive theoretical framework is needed for domain wall pinning in ferroelectrics with a coupling between polarization and lattice distortion, which would provide estimates of local coercivities, intrinsic domain wall widths as well as of correlation lengths for both phase and amplitude of the order parameter and their dependence on external field, in order to aid and guide quantitative interpretations of experimental results such as reported here.

The authors acknowledge helpful discussions with P. Chandra, S. Duttagupta, S. Ghosh, A. Krishnan, P.B. Littlewood, V.R. Palkar, R. Ramesh, J.F. Scott and M.M.J. Treacy. One of us (SB) especially acknowledges insightful discussions with A. I. Larkin about ferroelectric domain wall dynamics and its relation to other disordered elastic media.

Figure captions:

1. AFM images of (a) topography, (b) piezo-amplitude and (c) piezo phase signal of a 1μm x 1μm region which is poled in the down configuration after which a "cross" is written by a dc bias field which reverses the polarization of that spatial region. The domain wall separating up and down polarization has depleted amplitude, marked by the dark line in (b), consists of sharp and fuzzy regions. Panel (d) represents a composite image of topography and amplitude, which shows that the sharp domain wall precisely follows the contour of the contiguous grain boundaries. The "fuzzy" grain boundaries, two of which are encircled in each panel, have depleted amplitude over nearly the entire grain and the phase fluctuates randomly between 0 and $\pi$, seen in (c) in the form of fluctuating color.

2. Switching of a multigrain area shown by composite images of phase and topography of a 500nm x 500nm area at various polarizing dc voltages marked under each panel. Panel (a) represents the remanent state of the area after it was poled "down". Subsequent panels show the progressive switching of the region as the poling dc bias is increased. Most of the switching occurs via the motion of domain walls across contiguous grains, although independent nucleation also occurs, one of which is encircled in panel (c). The upper left corner in panel (a) shows a marked square area spanning a grain. The following panels show the progression of switching in this grain: first a sliver of the opposite phase enters and subsequently spreads across the grain in panels (c) through (i). A black arrow in the panels marks the moving front.

3. Single grain switching sequence in a faceted grain. The right columns are the phase images and the left columns are the amplitude images. The corresponding bias voltages are noted in the respective panels. Image area is 250nm x 250 nm. White dotted lines mark a few of the facets in panel (a). As the bias voltage increases, the back wall, marked by the white arrow, remains pinned while the front wall, marked by the black arrow, moves in a direction roughly parallel to the back wall, climbing a large coercivity-gradient that exists between the two walls. See text for discussions.

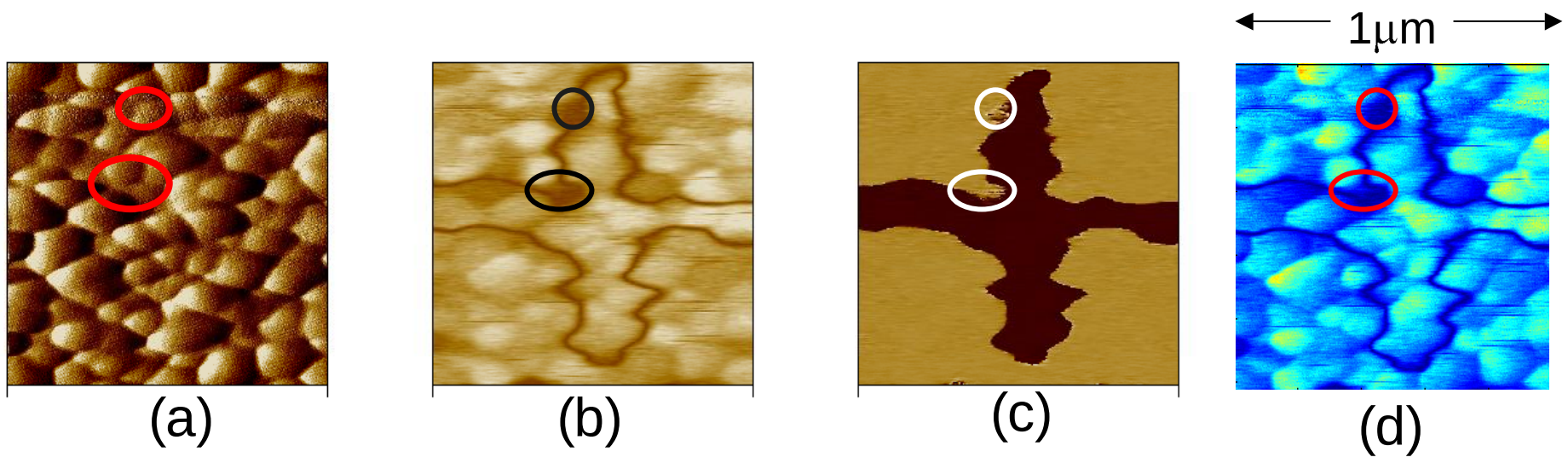

Figure 1

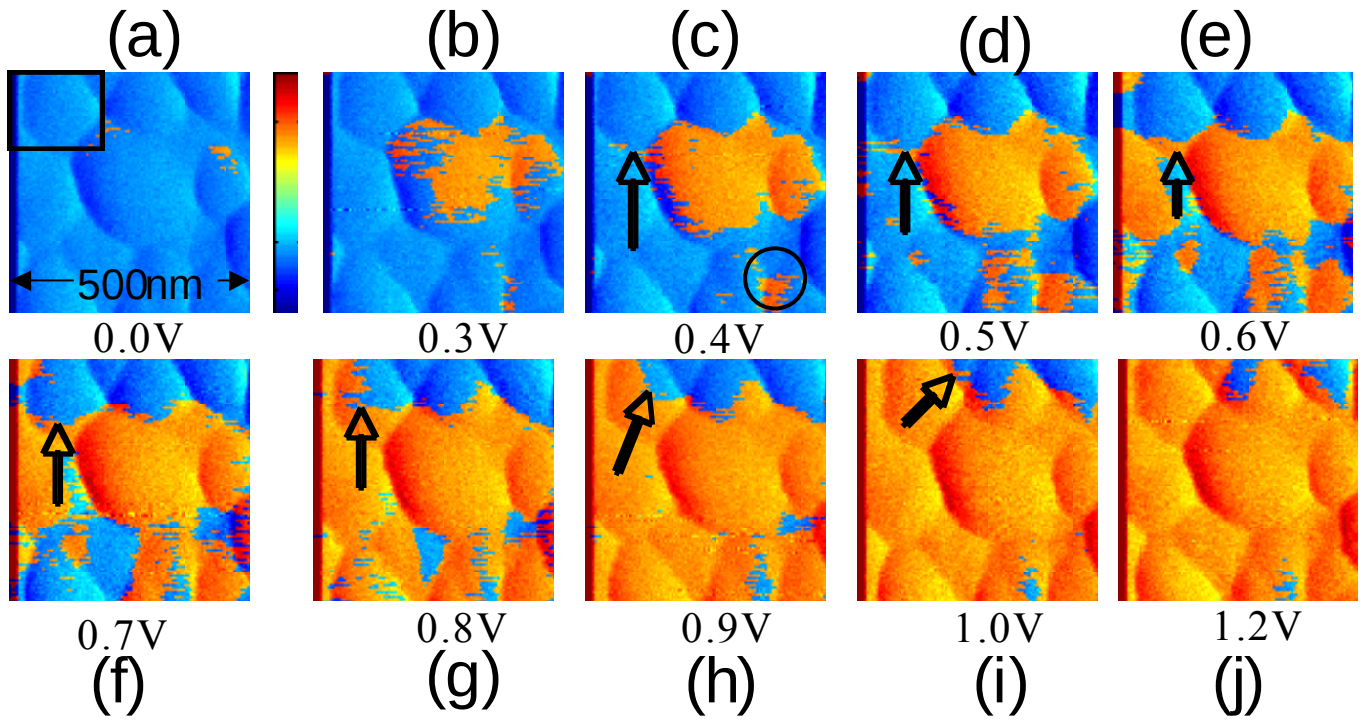

Figure 2

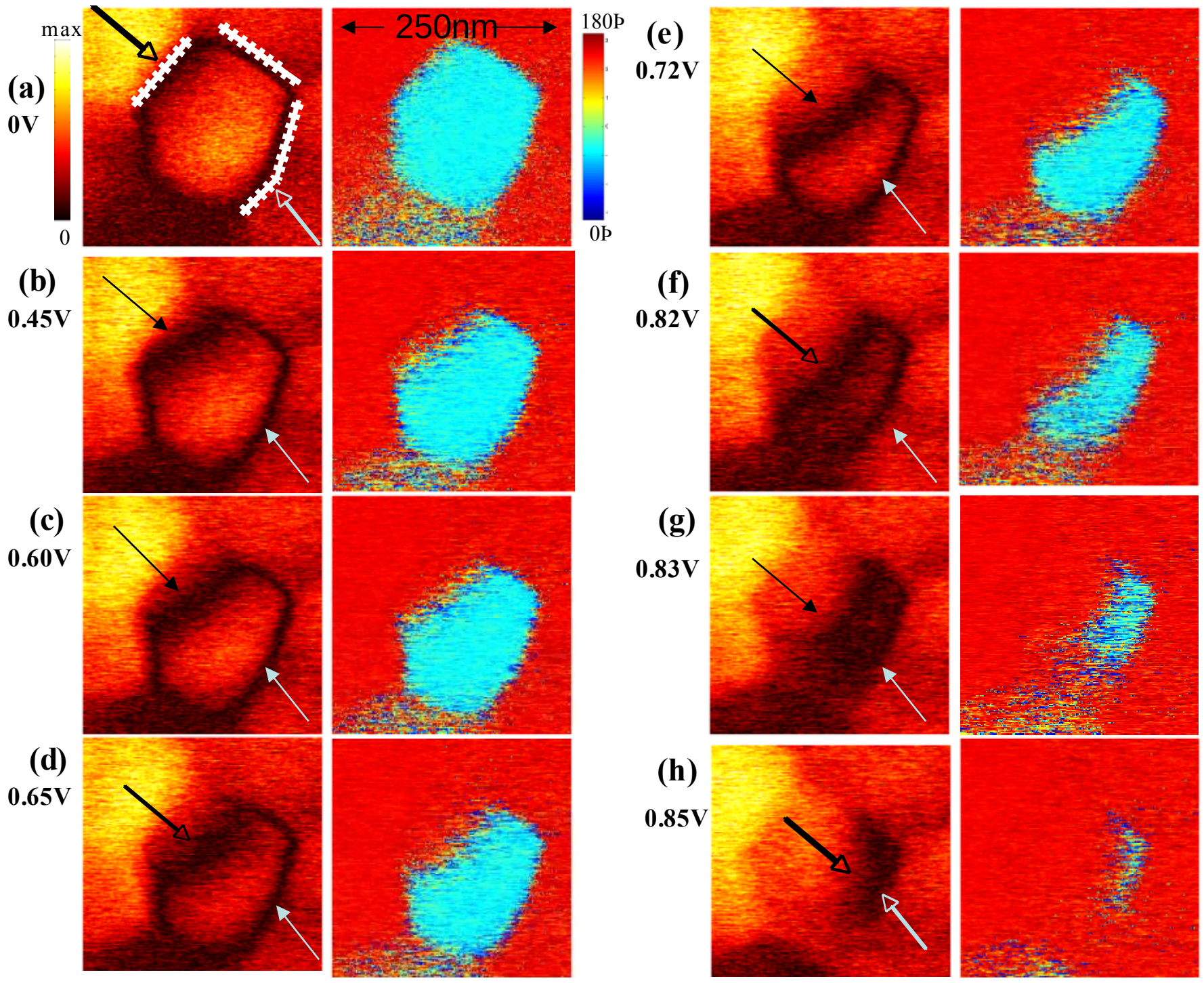

Figure 3